\documentclass[prb,twocolumn,showpacs]{revtex4}
\usepackage[dvips]{graphicx}
\usepackage{epsfig}
\usepackage{amsmath,amsfonts,amssymb,bm}
\usepackage{bm}
\usepackage[caption=false]{subfig}
\begin{document}
\title{Many-Body Delocalization in Strongly Disordered System with Long-Range Interactions: Finite Size Scaling}
\author{Alexander L. Burin}
\affiliation{Department of Chemistry, Tulane University, New
Orleans, LA 70118, USA}
\date{\today}
\begin{abstract}
Many-body localization in a disordered system of interacting spins coupled by the long-range interaction  $1/R^{\alpha}$ is investigated combining analytical theory considering resonant interactions and a finite size scaling of exact numerical solutions with a number of spins $N$.  The numerical results for a one-dimensional system are consistent with the general expectations of analytical theory for $d$-dimensional system including the absence of localization in the infinite system at $\alpha<2d$ and a universal scaling of a critical energy disordering $W_{c} \propto N^{\frac{2d-\alpha}{d}}$. 
\end{abstract}

\maketitle


\section{Introduction}

Localization - delocalization transition separates non-ergodic, reversible behavior from a chaotic, irreversible regime in quantum systems. Therefore it is in the focus of the scientific community since the concept of localization has been suggested for a single particle in a random field \cite{AndersonClassic}. At present localization in interacting disordered systems attracts growing  interest particularly because of its significance in quantum informatics \cite{PolkovnikovRevModPhys} where chaotic behavior can reduce a quantum hardware performance, and in atomic physics \cite{PolkovnikovRevModPhys,HusePhen2,PreprintML1,PreprintML2} where many-body systems can be constructed and studied using cold neutral atoms in a magneto-optical trap.  

A single particle model is relevant at temperature approaching zero where the number of excitations in the whole system is small so their interaction can be approximately neglected. At a finite temperature many-body interaction  complicates the localization problem because it can initiate the irreversible energy transport stimulating the particle transport \cite{FleishmanAnderson1980} and because of the Fock space complexity \cite{Levitov0}.  Many-body interaction can result in a single-particle localization breakdown for electrons in low dimensional  disordered metals ($d=1,2$) \cite{AronovReview,MirlinMBDel,Basko06}, quantum defects in quantum crystals \cite{KMlocalization85},  interacting two level systems in amorphous solids \cite{ab88Rv}  and molecular vibrations coupled by anharmonic interactions \cite{Leitner05}. 
A dramatic effect of the long-range many-body spin-spin interaction $R^{-\alpha}$ destroying the localization transition in the infinite system even at arbitrarily strong disordering for $\alpha<2d$ has been predicted.\cite{ab88Rv} The vanishing of localization in the infinite system for the system of interaction spins with the distance independent interaction has been also pointed out both analytically and numerically.\cite{Shepelyanskii}

The recent numerical investigations of many-body localization employing exact diagonalization methods \cite{HusePhen2,PreprintML2,OganesyanParamStudy,Reichmana,IoffeFeigelman,HuseNewOrderParameter,
HusePhen1,Pino}   have produced  a high resolution view of the localization transition. Finite size scaling of these numerical results strongly supports the existence of a many-body localization in  a strongly disordered one-dimensional system with a short-range interaction in the thermodynamic limit of an infinite system  (see also the rigorous proof in \cite{Math}).  The specific nature of the localization transition and its relationship to the previously developed analytical theories  \cite{Levitov0,Basko06,MirlinMBDel} remain unclear, particularly because of the absence of analytical dependencies that can be used to interpret numerical results. 
The investigation of a localization in a spin system with the long-range interactions \cite{PreprintML2} suggests the absence of localization in the infinite size limit for $1/R$ spin-spin interaction leaving the situation with interactions decreasing faster inconclusive.

The aim of the present work is the investigation of a many-body localization in systems with the long-range interaction decreasing with the distance $R$ as $R^{-\alpha}$  using the finite size scaling method. This method is very convenient to reveal the power law dependence of critical disordering on the number of spins $W_{c} \propto N^{\frac{2d-\alpha}{d}}$ (see Sec. II; this behavior contrasts to that for the short range interaction) and the lower constraint for the interaction law exponent $\alpha > 2d$ still permitting the localization in the infinite system predicted in  Ref. \cite{ab88Rv}. Moreover this and other scaling relationships (see Table \ref{tbl:scalinga}) absent in the case of a short-range interaction can serve here as guidelines for understanding of a very complicated many-body localization transition. 
It is worth to notice that the long-range interaction decreasing with the distance according to the power law inevitably exists between quasiparticles possessing either charges or dipolar, magnetic or elastic moments and it can significantly influence the localization transition similarly to a single particle case \cite{AndersonClassic,Levitov1} or even stronger \cite{ab88Rv}.

Below in Sec. \ref{sec:Analyt} we briefly introduce the analytical theory of a many-body localization breakdown due to the energy delocalization in an ensemble of interacting resonant pairs (Fig. \ref{fig:ResPairs}) for the arbitrarily system dimension $d$. The dependence of the localization threshold disordering $W_{c}$ on the system size is derived. Then in Sec. \ref{sec:FinSizeSc} the numerical results for localization threshold at $d=1$ are described and compared to the analytical theory.

\section{Analytical approach}
\label{sec:Analyt}

We investigate the model of $N$ interacting spins $1/2$ placed  into equally spaced sites of a $d$-dimensional hypercube with the spatial density $n$. These spins are subjects to uncorrelated  random $z-$directional fields ($\phi_{i}S_{i}^{z}$) uniformly distributed within the domain $(-W/2, W/2)$. The interaction between spins $i$ and $j$ has two components $U_{ij}S_{i}^{z}S_{j}^{z}$ and $V_{ij}(S_{i}^{+}S_{j}^{-} + S_{i}^{-}S_{j}^{+})$ depending on  the distance as $u_{ij}/R_{ij}^{\beta}$, $v_{ij}/R_{ij}^{\alpha}$ with $\alpha=\beta$ and random sign interaction constants $u_{ij}, v_{ij} = \pm U_{0}$ as in the ``anisotropic'' interaction case.\cite{PreprintML2}.  The more general case $\alpha \neq \beta$ is interesting but more complicated  \cite{PreprintML2} and requires a special consideration. 
Our consideration is restricted to a relatively large interaction exponents $\alpha \geq d$ so Anderson localization of single particle excitations is possible in the infinite size limit.

\begin{figure}[h!]
\centering
\includegraphics[width=3cm]{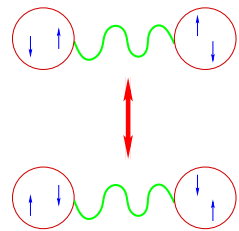}
\caption{Two resonant pairs coupled by the many-body spin-spin interaction and their joint transition.}
\label{fig:ResPairs}
\end{figure}

In a strongly disordered system where the interaction at the average distance is much smaller than disordering, $\tilde{U}=U_{0}n^{\frac{\alpha}{d}}\ll W$,  spin dynamics takes place in sparse resonant pairs where the change in the diagonal (Ising) energy  \cite{PreprintML2} $\mid\Delta_{ij}\mid=\mid \phi_{i}-\phi_{j}+\sum_{k\neq i,j}(U_{ik}-U_{jk})S^{z}_{k}\mid$ due to the flip-flop transition of spins $i$ and $j$ is smaller than the flip-flop interaction $V_{ij}$. Two spins separated by the distance $R$ form the resonant pair with the probability $P(R)\approx \frac{U_{0}}{WR^{\alpha}}$ (cf. Refs. \cite{ab88Rv,Levitov1}). Consequently 
the density of resonant pairs of a certain size $R$ (second spin can occupy the volume $R^d$) can be estimated as 
\begin{eqnarray}
n_{p}(R) \sim nR^{d}P(R)=
n\frac{\tilde{U}}{W}
\left(\frac{1}{nR^{d}}\right)^{\frac{\alpha-d}{d}}. 
\label{eq:respairs}
\end{eqnarray}
Then the characteristic flip-flop interaction of resonant pairs (see Fig. \ref{fig:ResPairs}) can be expressed using their  interaction at the average distance between them, $n_{p}^{-\frac{1}{d}}$, as 
\begin{eqnarray}
V(R) \sim U_{0}n_{p}(R)^{\frac{\alpha}{d}}=\tilde{U}\left(\frac{\tilde{U}}{W}\right)^{\frac{\alpha}{d}}\left(\frac{1}{nR^{d}}\right)^{\frac{\alpha(\alpha-d)}{d^2}}. 
\label{eq:flflresp}
\end{eqnarray}
The energy delocalization within the subset of pairs of a certain size $R$ is expected if their coupling $V(R)$ exceeds 
the characteristic energy (disordering) of such pairs of the size $R$ given by 
\begin{eqnarray}
E(R) \sim \frac{U_{0}}{R^{\alpha}}=\tilde{U}\frac{1}{\left(nR^{d}\right)^{\frac{\alpha}{d}}}. 
\label{eq:PairEnergy}
\end{eqnarray}

Comparing behaviors of coupling strengths Eq. (\ref{eq:flflresp}) and typical energies Eq. (\ref{eq:PairEnergy}) one can see  that at sufficiently small interaction exponent  $\frac{\alpha(\alpha-d)}{d^2}<\frac{\alpha}{d}$ ($\alpha<2d$) the flip-flop interaction Eq. (\ref{eq:flflresp}) always exceeds the typical energy of pairs Eq. (\ref{eq:PairEnergy}) at sufficiently large $R$. {\it Consequently the  delocalization should take place at arbitrarily disordering for sufficiently large system size if $\alpha<2d$.}
The delocalization within the subsystem of resonant pairs forms a dense ergodic subsystem that should serve as a thermal bath for the rest of spins leading to the  irreversible ergodic dynamics of the whole system \cite{ab88Rv}.

In the finite ensembles of interacting spins relevant for quantum informatics and cold atomic systems \cite{PolkovnikovRevModPhys,HusePhen2,PreprintML1,PreprintML2} the delocalization takes place at $\alpha < \alpha_{*}=2d$  starting with the system size $R=R_{*}\approx n^{-\frac{1}{d}}\left(\frac{W}{\tilde{U}}\right)^{\frac{1}{2d-\alpha}}$ where the flip-flop interaction of resonant pairs Eq. (\ref{eq:flflresp}) approaches their typical energy Eq. (\ref{eq:PairEnergy}) ($V(R_{*})=E(R_{*})$). Consequently one can express the system size or spin number dependence of critical disordering $W_{c}$ separating localization ($W>W_{c}$) and delocalization ($W<W_{c}$) domains  as 
\begin{eqnarray}
W_{c} \approx \tilde{U} (nR^d)^{\frac{2d-\alpha}{d}}=\tilde{U} N^{\frac{2d-\alpha}{d}}. 
\label{eq:MainScaling}
\end{eqnarray}
The threshold case of $\alpha=2d$  where the number of resonant interactions increases logarithmically with the system size needs special study (cf. Ref.  \cite{Levitov1}). 

The prediction of Eq. (\ref{eq:MainScaling}) for the size dependent localization threshold is consistent with the analytical and numerical investigations of many-body localization transition for the distance independent interactions ($\alpha=0$, $W_{c} \propto \tilde{U}N^{2}$)\cite{Shepelyanskii} though the model investigated there is not fully identical to the present model. One should notice that the coupling strength of spins is rescaled in that model by the factor of $N^{-\frac{1}{2}}$.

The power law dependence Eq. (\ref{eq:MainScaling}) will be used below in the numerical finite size scaling analysis of many-body localization and the logarithmic dependence of $W_{c}(N)$ will  serve as a natural crossover between the unlimited increase of $W_{c}(N)$ or its saturation at different exponents $\alpha$ in  the infinite size limit. 
 The results for the predicted critical behaviors are summarized in Table \ref{tbl:scalinga}. 


\begin{table}
 \caption{Predictions for the critical interaction exponent $\alpha_{*}$, critical number of spins $N_{*}$ and system size $R_{*}$ at a given disordering $W$ and critical disordering $W_{*}$ at a given number of spins $N$ where delocalization takes place in the system with the long-range spin-spin interactions $r^{-\alpha}$ for $\alpha<\alpha_{*}$. 
 } 
 \label{tbl:scalinga}
\centering
\begin{tabular}{|l|c|c|r|}
  \hline
   $\alpha_{*}$ & $N_{*}$ ($\alpha<\alpha_{*}$) & $R_{*}$ ($\alpha<\alpha_{*}$) & $W_{*}$ ($\alpha<\alpha_{*}$)  \\
   \hline
   $2d$ & $\left(\frac{W}{\tilde{V}}\right)^{\frac{d}{2d-\alpha}}$ & $n^{-\frac{1}{d}}\left(\frac{W}{\tilde{V}}\right)^{\frac{1}{2d-\alpha}}$ & $\tilde{U}N^{\frac{2d-\alpha}{d}}$ \\ 
   \hline  
  \end{tabular}
\end{table}



\section{Finite size scaling}
\label{sec:FinSizeSc}

The numerical calculations in the present work are somewhat similar and somewhat different from that of Ref. \cite{PreprintML2} We analyze the infinite time spin-spin correlation function, which is similar to the dynamic polarization studied there and also the level statistics which can be used to characterize the localization-delocalization transition. \cite{ShklShapiro} Also we collect statistics not from all systems eigenstates,\cite{PreprintML2} but from only states having energies close to zero, where the density of states approaches maximum. Although the results should not be different in an infinite system limit we expect that the finite size effect will be smaller in our consideration because of the excluded contributions of the localized states at the edges of the spectrum.

\begin{figure}[h!]
\centering
\subfloat[]{\includegraphics[scale=0.3]{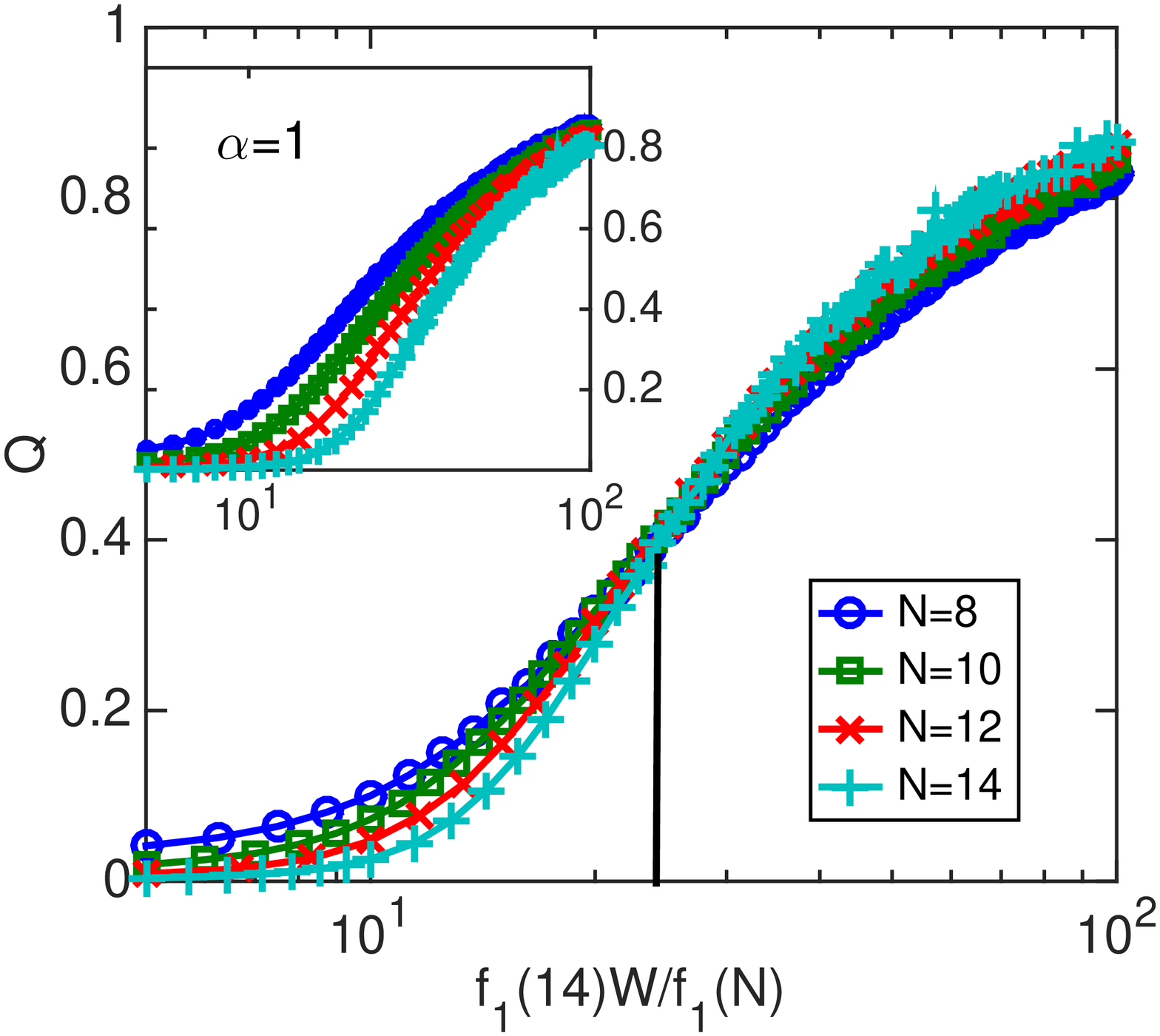}}\\
\subfloat[]{\includegraphics[scale=0.3]{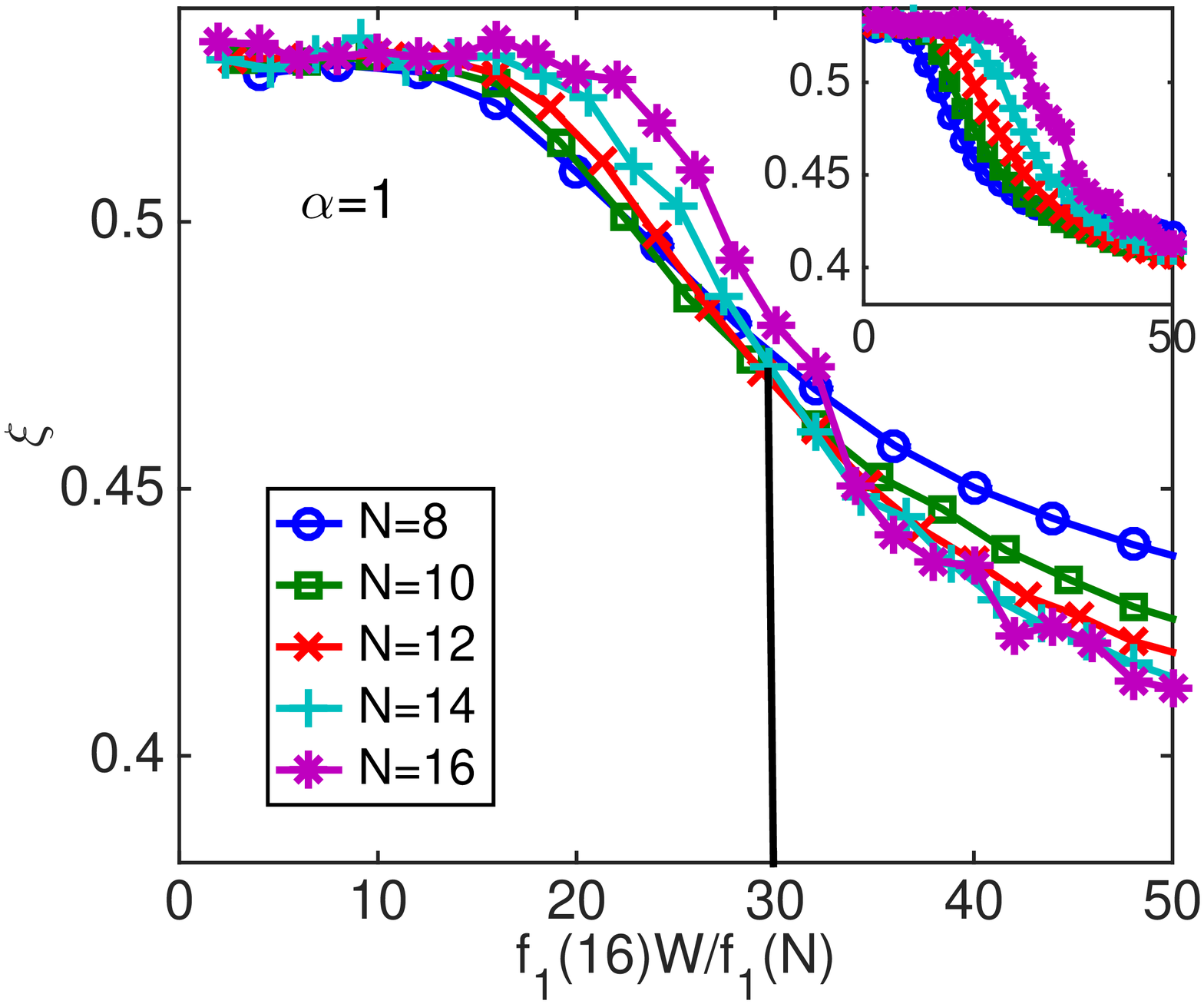}}\\
{\caption{\small  Rescaled  dependence of ergodicity (a) and level statistics (b) parameters on disordering $W$ compared to the original dependence (insets) for $R^{-1}$ spin-spin interactions. Here and in other figures vertical straight lines show the estimate for the localization transition point }
\label{fig:a1}}
\end{figure} 

Also in addition to the qualitative visual data analysis \cite{PreprintML2} we develop the quantitative scaling analysis  of data which permits us to extract the size dependence of the localization threshold and compare it with the predictions of analytical model (Table \ref{tbl:scalinga}). Unfortunately, any numerical finite size scaling for the many body localization problem  is limited to a very small system size because of the exponential increase in the number of states with this size. Yet the power law scaling of key system parameters in the case of power law interactions can be easier identified numerically then the unclear behavior in the systems with the short-range interactions. This is seen from the analysis below  demonstrating the localization transition scaling for interactions decreasing with the distance slower than $R^{-2d}$. Therefore we believe that the proposed scaling method works better for systems under consideration than in the case of a short range interaction and it can be help to understand the nature of many-body localization. 

The pure power law $N^{a}$ dependence of $W_{c}(N)$ is dramatically sensitive to the finite size effect  at spin numbers $N\sim 16$ and interaction exponents $\alpha$ close to the threshold $\alpha=2d$. At these sizes the logarithmic ($\ln(N)$) dependence fits well to $N^{0.35}$ power law. To account better for the finite size effects we used the scaling function  in the form
\begin{eqnarray}
f_{a}(N)=\frac{N}{2(N-1)}\left(2\sum_{n=1}^{N/2-1}n^{a-1}+\left(\frac{2}{N}\right)^{a-1}\right). 
\label{eq:sc_func}
\end{eqnarray} 
This function represents the size dependence for the number of resonant interactions per spin decreasing with the distance as $n^{a-1}$, which is expected from the discrete version of the analytical derivation.\cite{ab88Rv} The prefactor $\frac{N/2}{N-1}$ accounts for the number of spins oriented in the opposite direction to the given spin so they can perform a joint flip-flop transition. 

In the large $N$ limit Eq. (\ref{eq:sc_func}) yields $f_{a}(N)\approx \frac{N^{a}}{a2^a}$ for $a>0$,  $f_{a}(N)\approx \ln(N)$ for $a=0$, while for $a<0$ it approaches the finite limit as $f_{a}(N)\approx \zeta(a-1)-\frac{1}{|a|2^a N^{|a|}}$, where $\zeta(x)$ stands for the Riemann zeta function. According to the analytical theory Eq. (\ref{eq:MainScaling}) one should expect $a=2-\alpha$ for the small power law exponent $\alpha<2$. If the same delocalization mechanism is applicable to $\alpha>2$ the negative exponent $a=2-\alpha$ asymptotic of Eq. (\ref{eq:sc_func}) can be approximately relevant for the threshold dependence on the system size, though the threshold approaches the finite value in the infinite size limit.  

\begin{figure}[h!]
\centering
\subfloat[]{\includegraphics[scale=0.3]{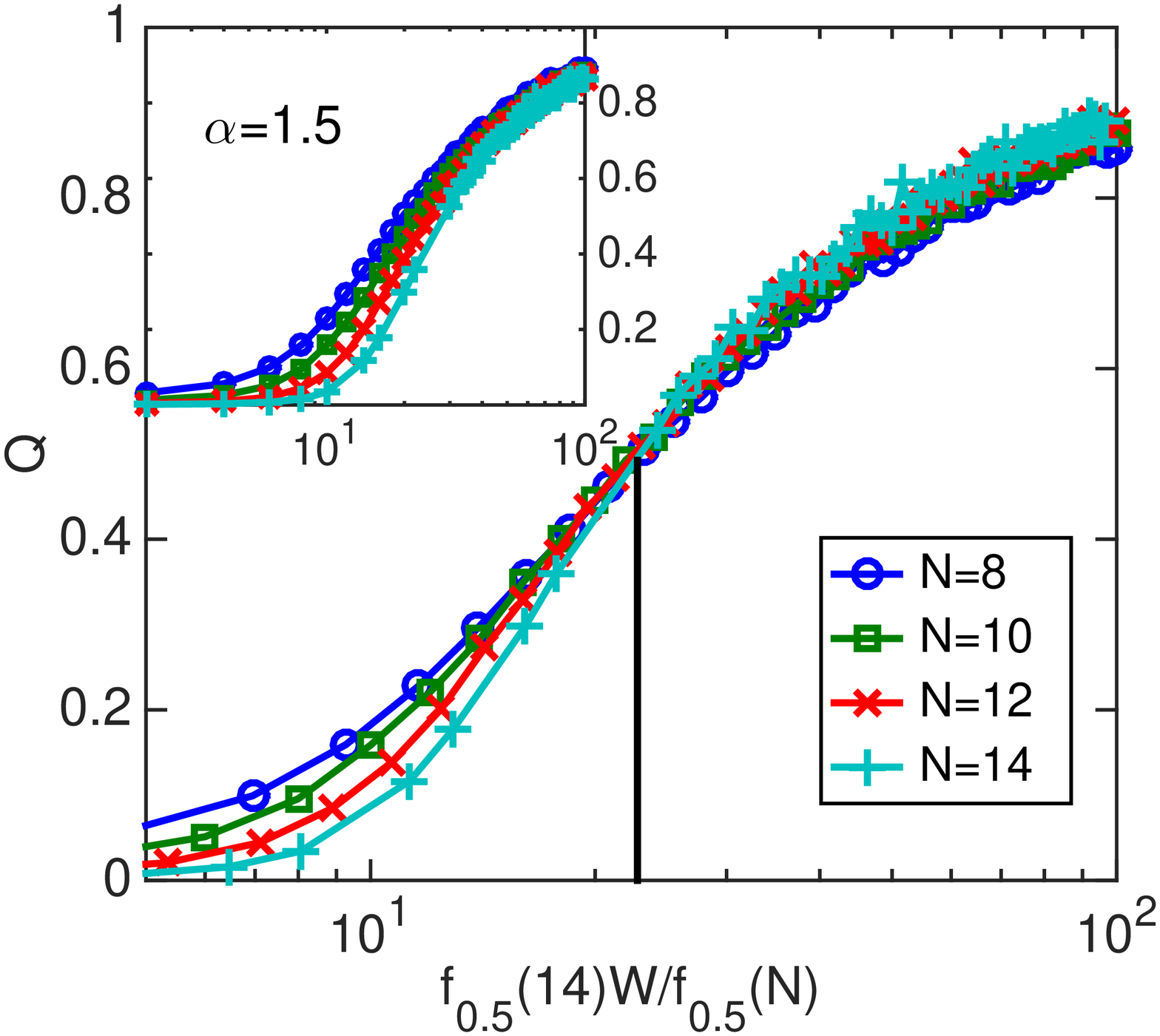}}\\
\subfloat[]{\includegraphics[scale=0.3]{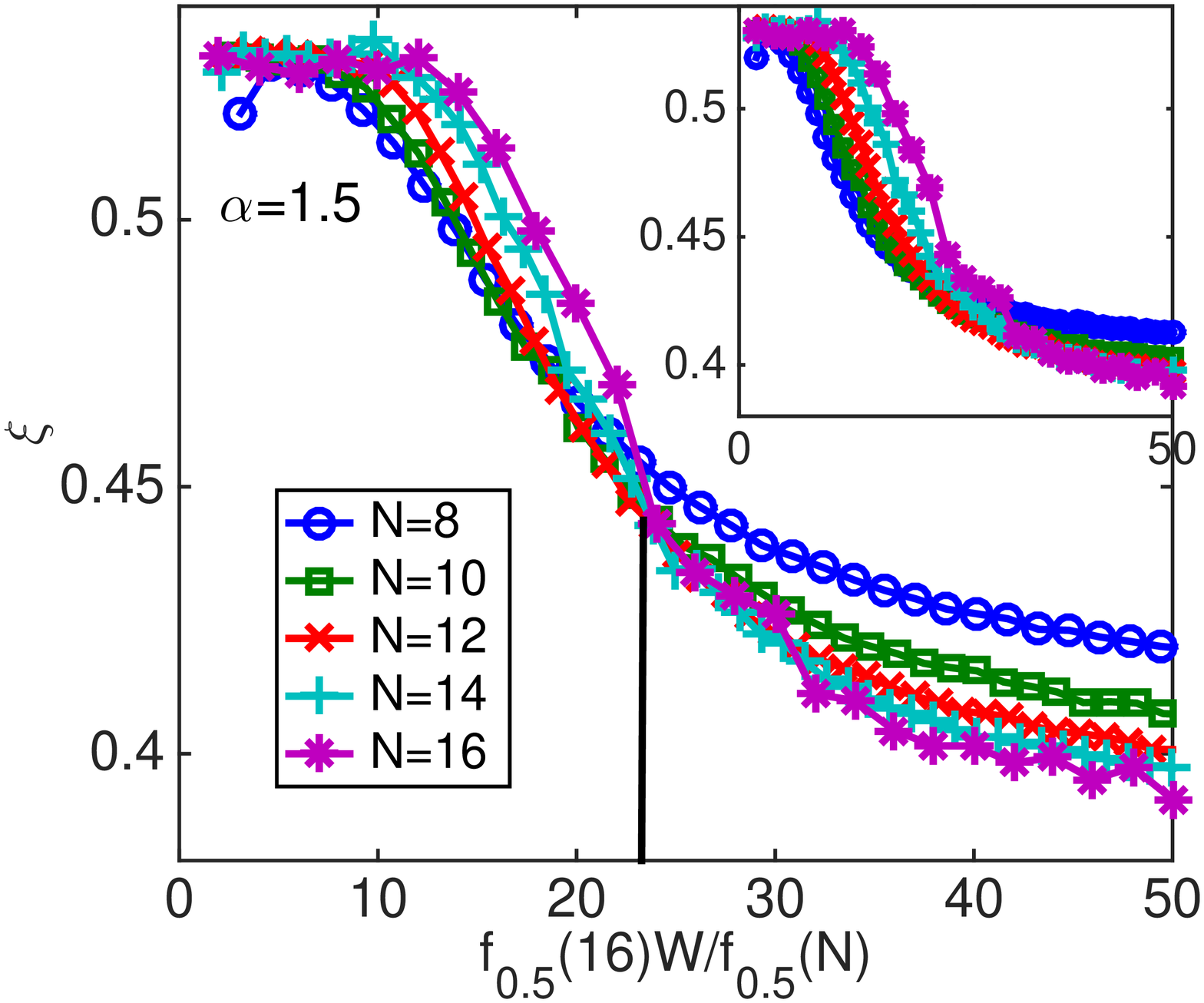}}\\
{\caption{\small  Rescaled  dependence of ergodicity (a) and level statistics (b) parameters on  disordering $W$ compared to the original dependence (insets) for $R^{-1.5}$ spin-spin interactions..}
\label{fig:a1_5}}
\end{figure} 

In the numerical analysis of a $1D$ system we set $U_{0}=n=1$ expressing disordering in $\tilde{U}$ units. Interspin distance $R_{ij}$ is taken as a minimum distance alone the closed chain ($R_{ij}=\min(|i-j|, N-|i-j|)$). 

\begin{figure}[h!]
\centering
\subfloat[]{\includegraphics[scale=0.3]{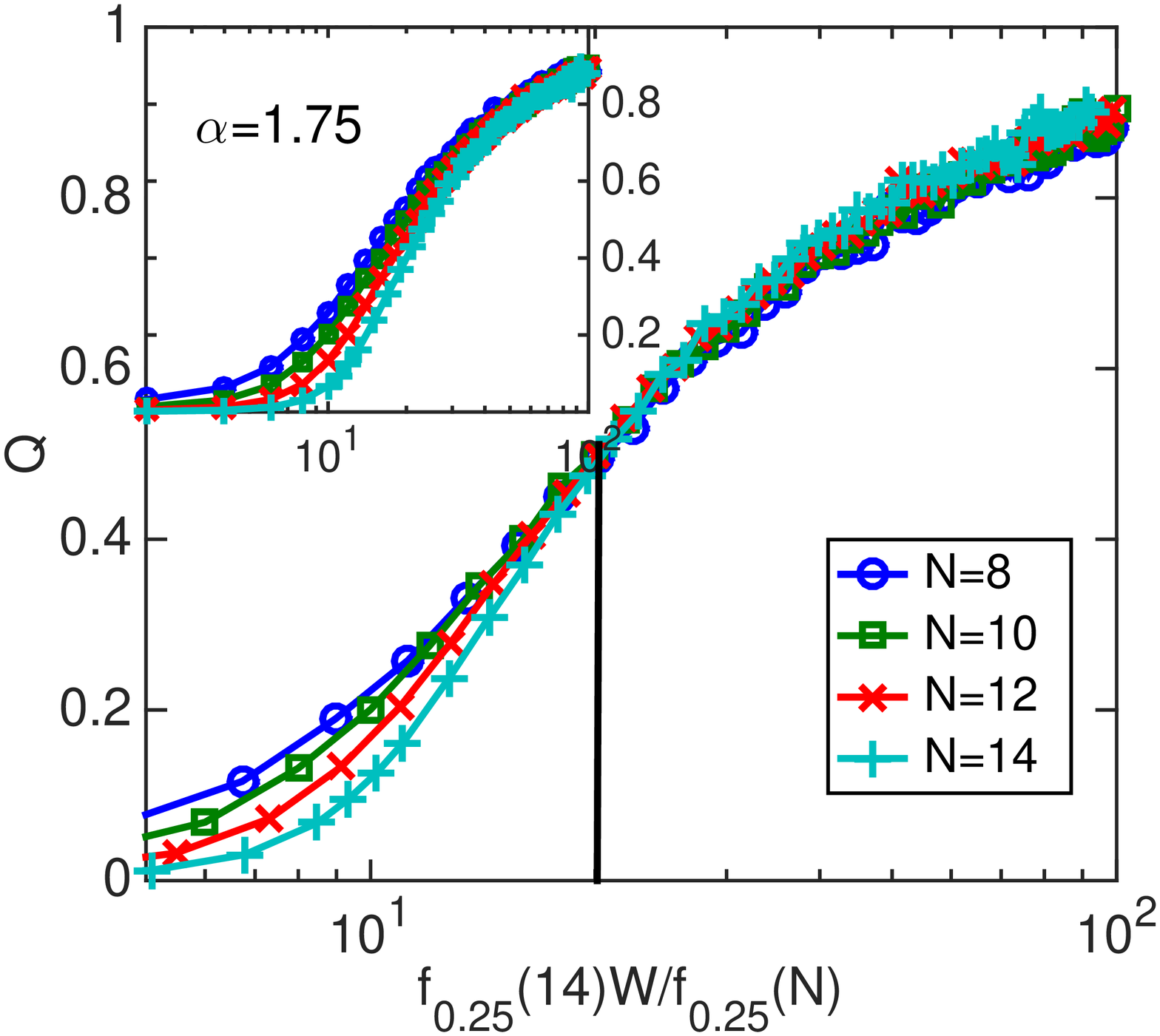}}\\
\subfloat[]{\includegraphics[scale=0.3]{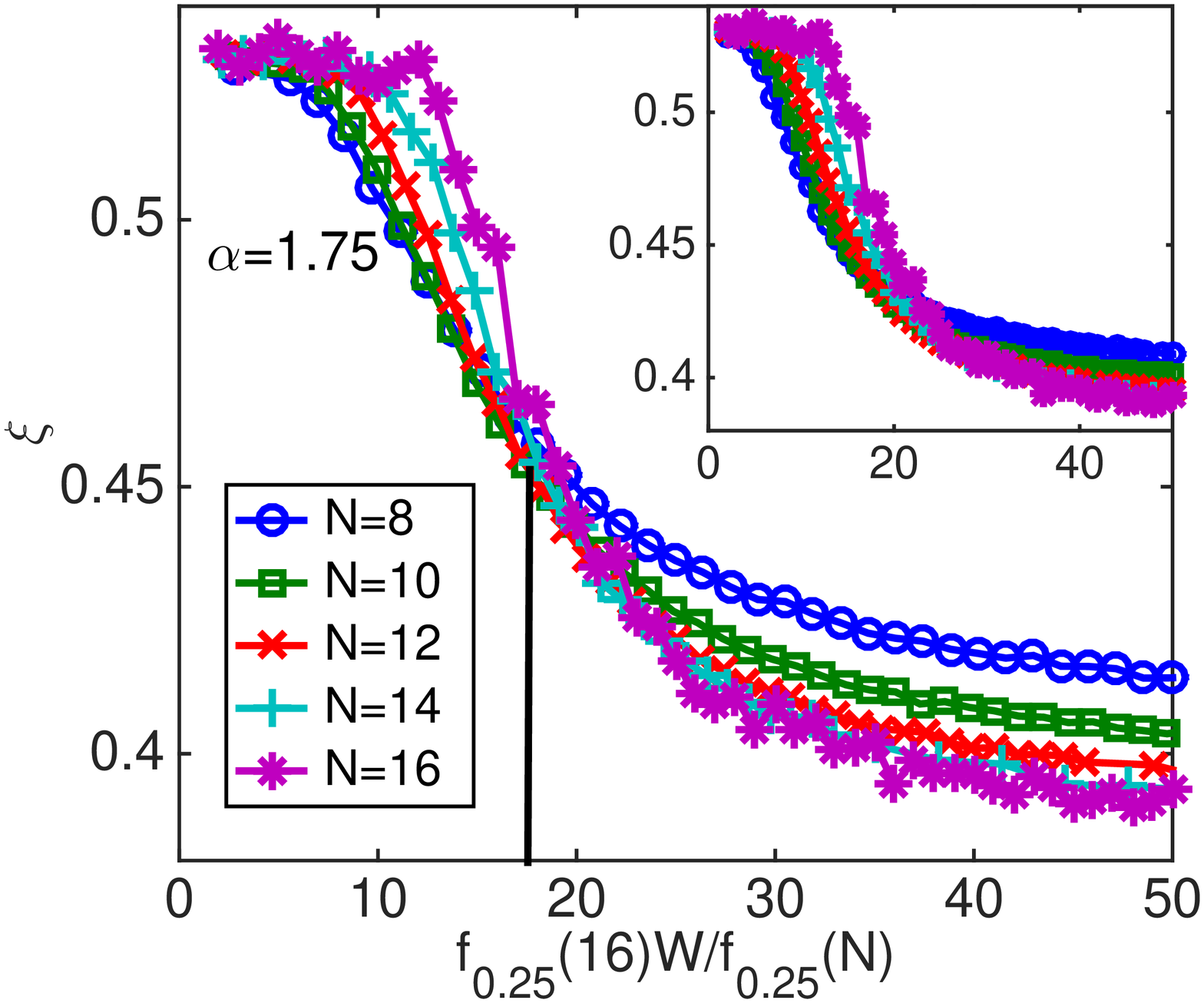}}\\
{\caption{\small  Rescaled  dependence of ergodicity (a) and level statistics (b) parameters on disordering $W$ compared to the original dependence (insets) for $R^{-1.75}$ spin-spin interactions..}
\label{fig:a1_75}}
\end{figure} 

To characterize the localization we use the ergodicity parameter defined as a configuration averaged spin-spin correlation function at infinite time\cite{BGold} 
\begin{eqnarray}
Q = 4<\delta S^{z}(\infty)\delta S^{z}(0)>'=\frac{1}{N_{\alpha}}\sum'_{\alpha}\mid <\alpha|\delta S^{z}|\alpha>|^2,
\nonumber\\ 
\delta S^{z}=S^{z}-<S^{z}>, 
\end{eqnarray}
The averaging $<...>'$ is performed  over the narrow band of eigenstates $\alpha$ around zero energy $(-\delta <E_{\alpha}< \delta)$, $\delta =0.04W\sqrt{N}$ and $N_{\alpha}$ is the number of states in this energy domain. The many-body density of states $g(E) \approx \exp\left(\frac{E^2}{24NW^2}\right)/\sqrt{24\pi NW^2}$ changes at the scale $\delta$ by around $1\%$ which is the reason for our choice of the bandwidth. It is estimated using the random potential part of the system Hamiltonian $-\sum_{k=1}^{N}\phi_{k}S_{k}^{z}$ assuming the law of large numbers ($N \gg 1$) and strong disordering $\tilde{U}\ll W$. The consideration of energies near $E=0$ corresponding to the maximum density of states is approximately equivalent to the infinite temperature limit which is found very convenient to characterize many-body localization transition. \cite{PreprintML2,OganesyanParamStudy,Reichmana}

The ergodicity parameter $Q$ approaches zero in delocalization regime ($N\rightarrow \infty$) and remains finite otherwise. We also consider  the level statistics \cite{OganesyanParamStudy} characterized by the average ratio of the minimum of two subsequent energy differences of adjacent states and the maximum of those two differences $\xi=<min(\delta E_{i}, \delta E_{i+1})/min(\delta E_{i}, \delta E_{i+1})>$ ($\delta E_{i}=E_{i+1}-E_{i}$). The parameter $\xi$ approaches its maximum $0.53$ in the delocalization regime (Wigner-Dyson level statistics,\cite{ShklShapiro}) while in the strong localization regime (Poisson statistics) it has the minimum $\xi \approx 0.38$.\cite{OganesyanParamStudy}

\begin{figure}[h!]
\centering
\subfloat[]{\includegraphics[scale=0.3]{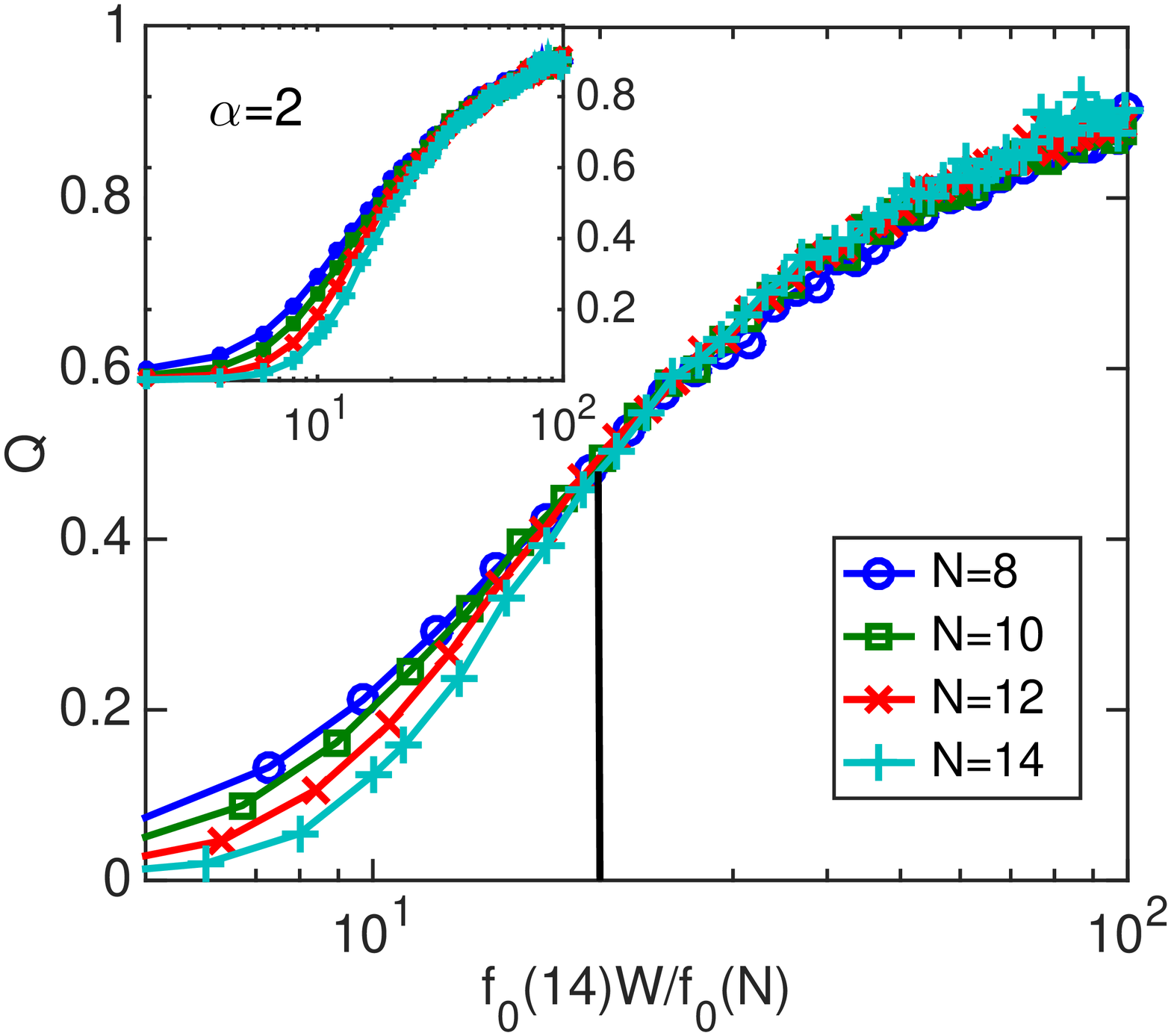}}\\
\subfloat[]{\includegraphics[scale=0.3]{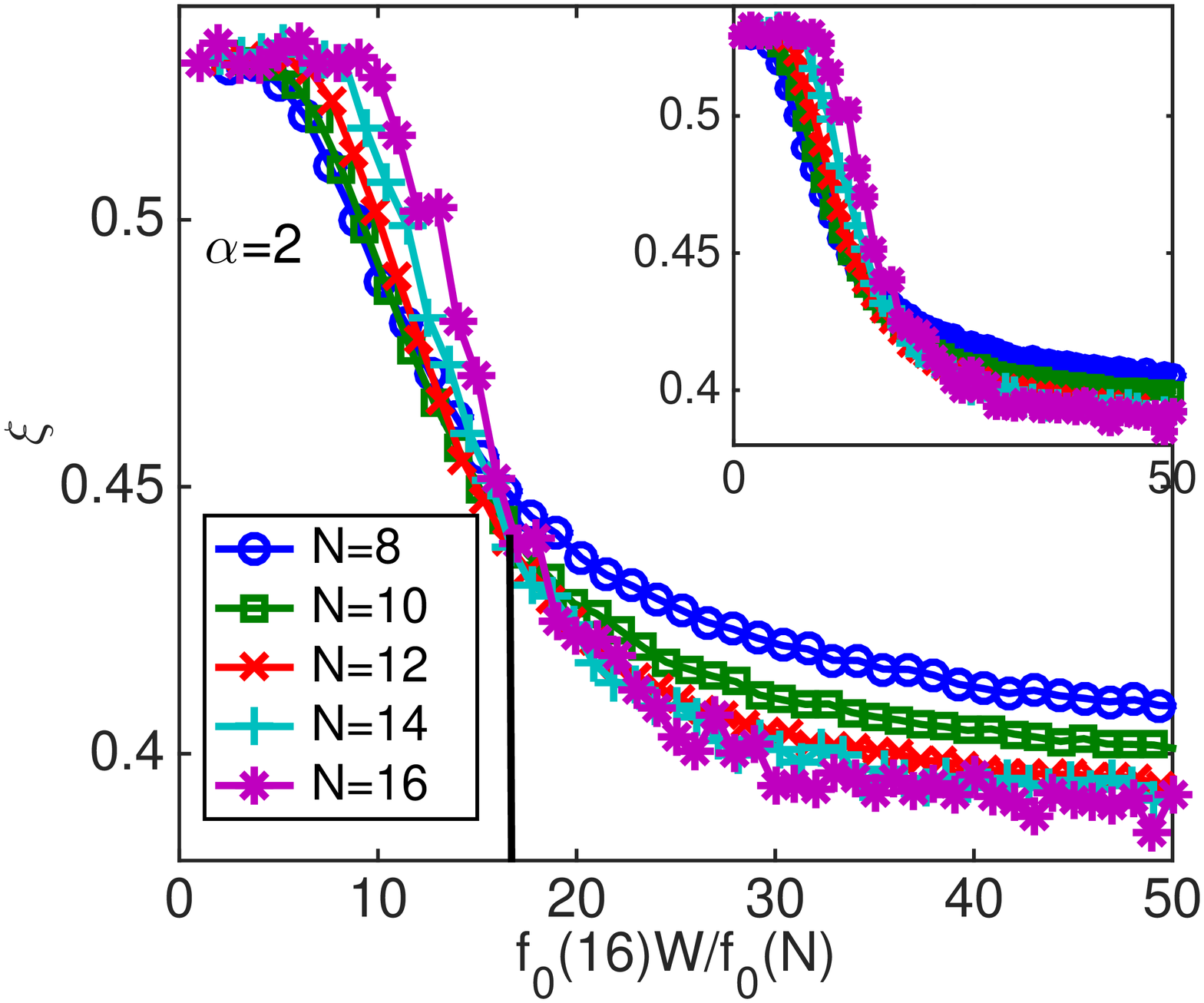}}\\
{\caption{\small  Rescaled  dependence of ergodicity (a) and level statistics (b) parameters on  disordering $W$ compared to the original dependence (insets) for $R^{-2}$ spin-spin interactions..}
\label{fig:a2}}
\end{figure}

\begin{figure}[h!]
\centering
\subfloat[]{\includegraphics[scale=0.3]{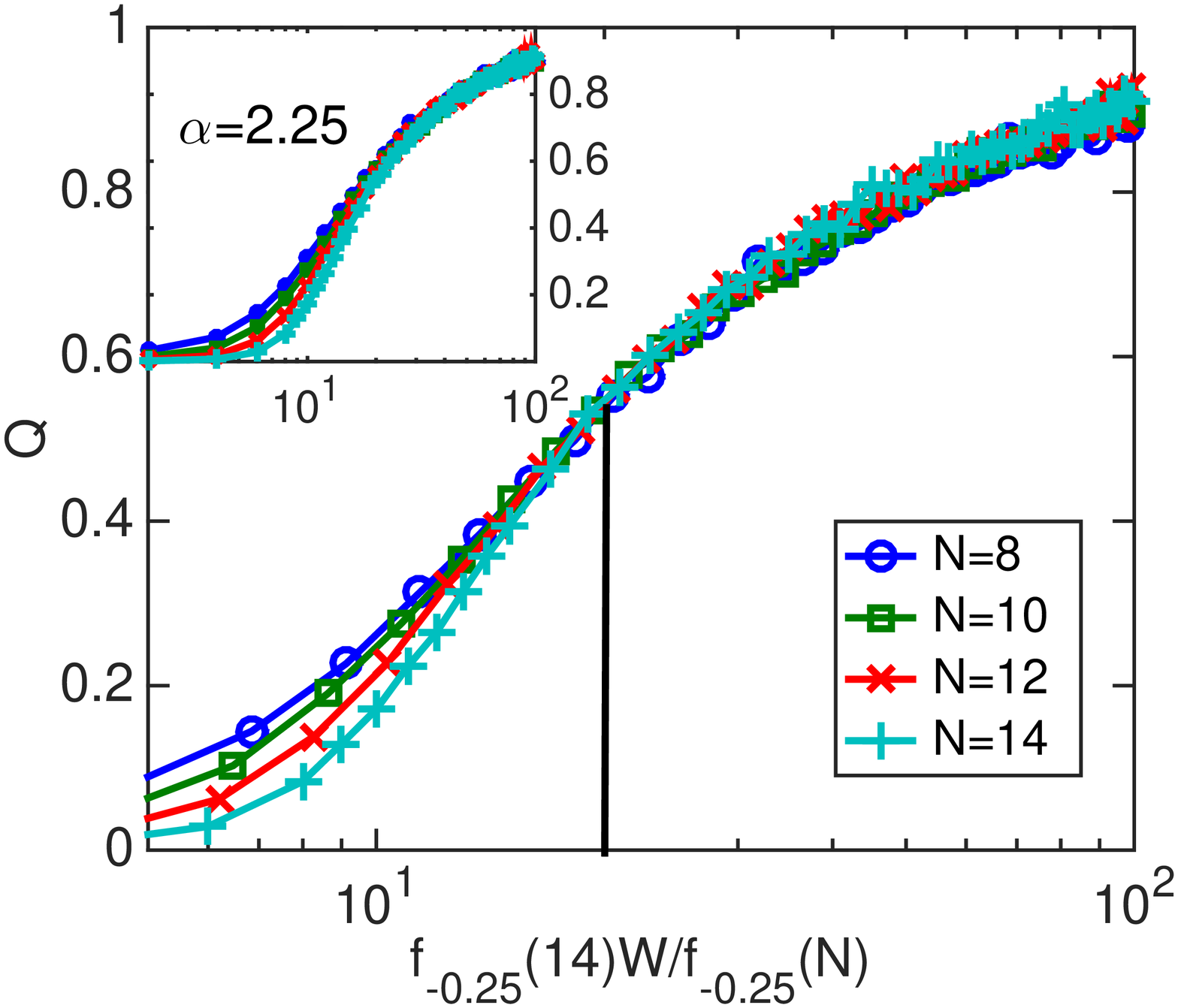}}\\
\subfloat[]{\includegraphics[scale=0.3]{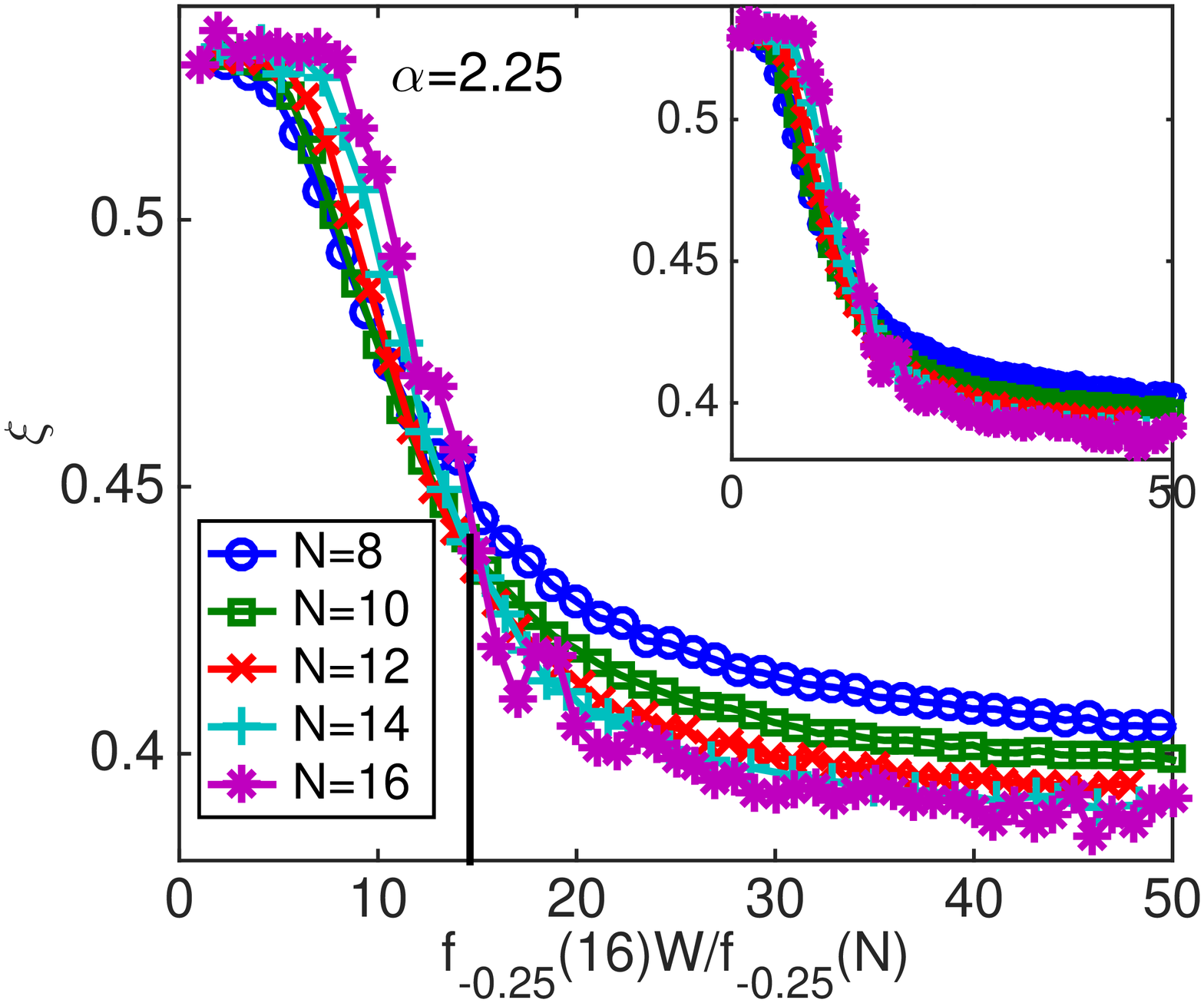}}\\
{\caption{\small  Rescaled  dependence of ergodicity (a) and level statistics (b) parameters on disordering $W$ compared to the original dependence (insets) for $R^{-2.25}$ spin-spin interactions..}
\label{fig:a2_25}}
\end{figure} 

Calculations of ergodicity parameter and level statistics were performed using Matlab software. 
Random Hamiltonians have been generated for interaction exponents $\alpha=\beta$ ranging between $1$ and $3$, disordering $2< W <100$, and total even numbers of spins $8 \leq N \leq 16$. The conserving projection of  the total spin to the $z$-axis has been always set to $0$ in agreement with the assumption of an infinite temperature. The results have been averaged over a  sufficiently large number of realizations chosen to make the relative error of the estimate less than $1\%$. 

\begin{figure}[h!]
\centering
\subfloat[]{\includegraphics[scale=0.3]{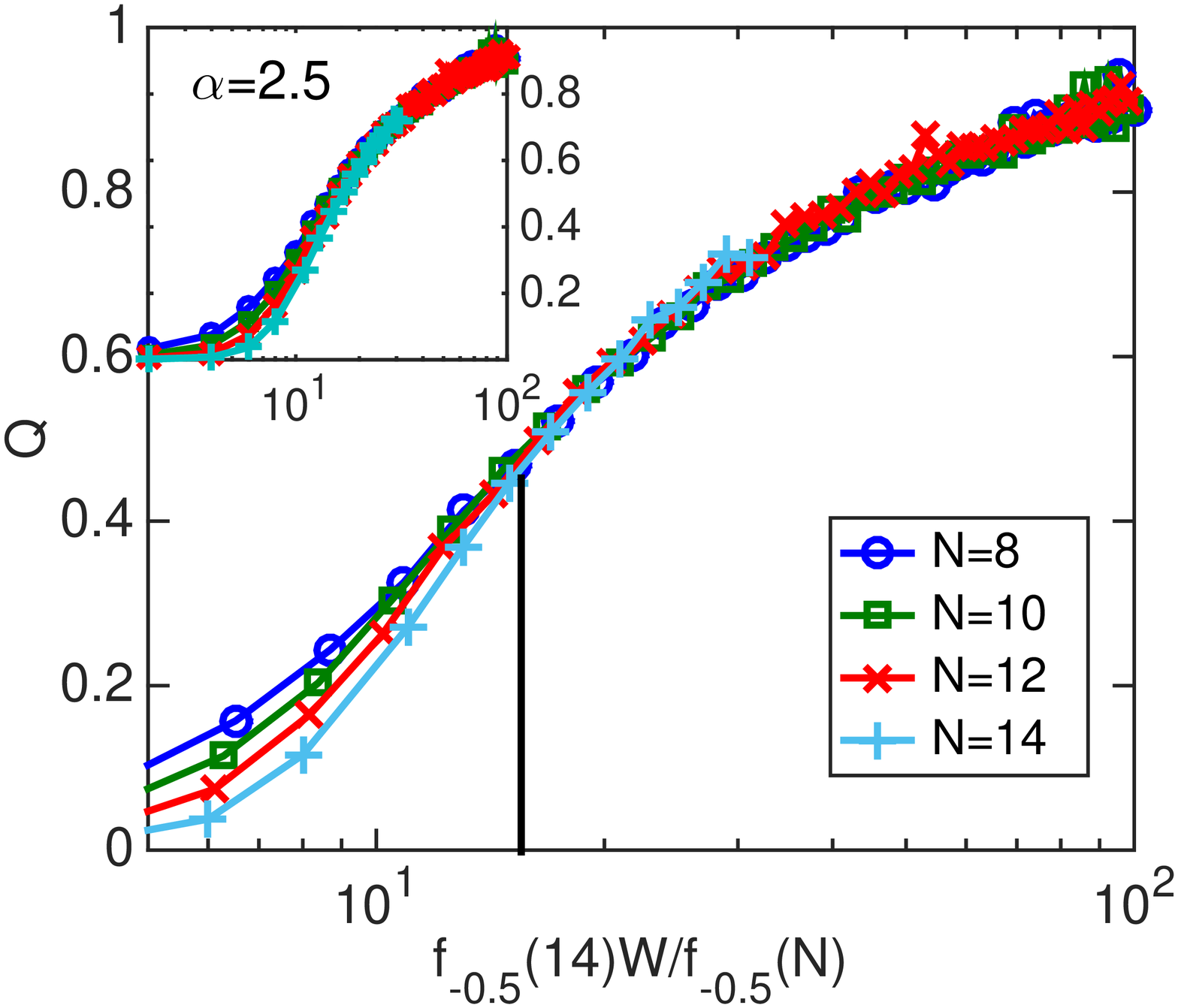}}\\
\subfloat[]{\includegraphics[scale=0.3]{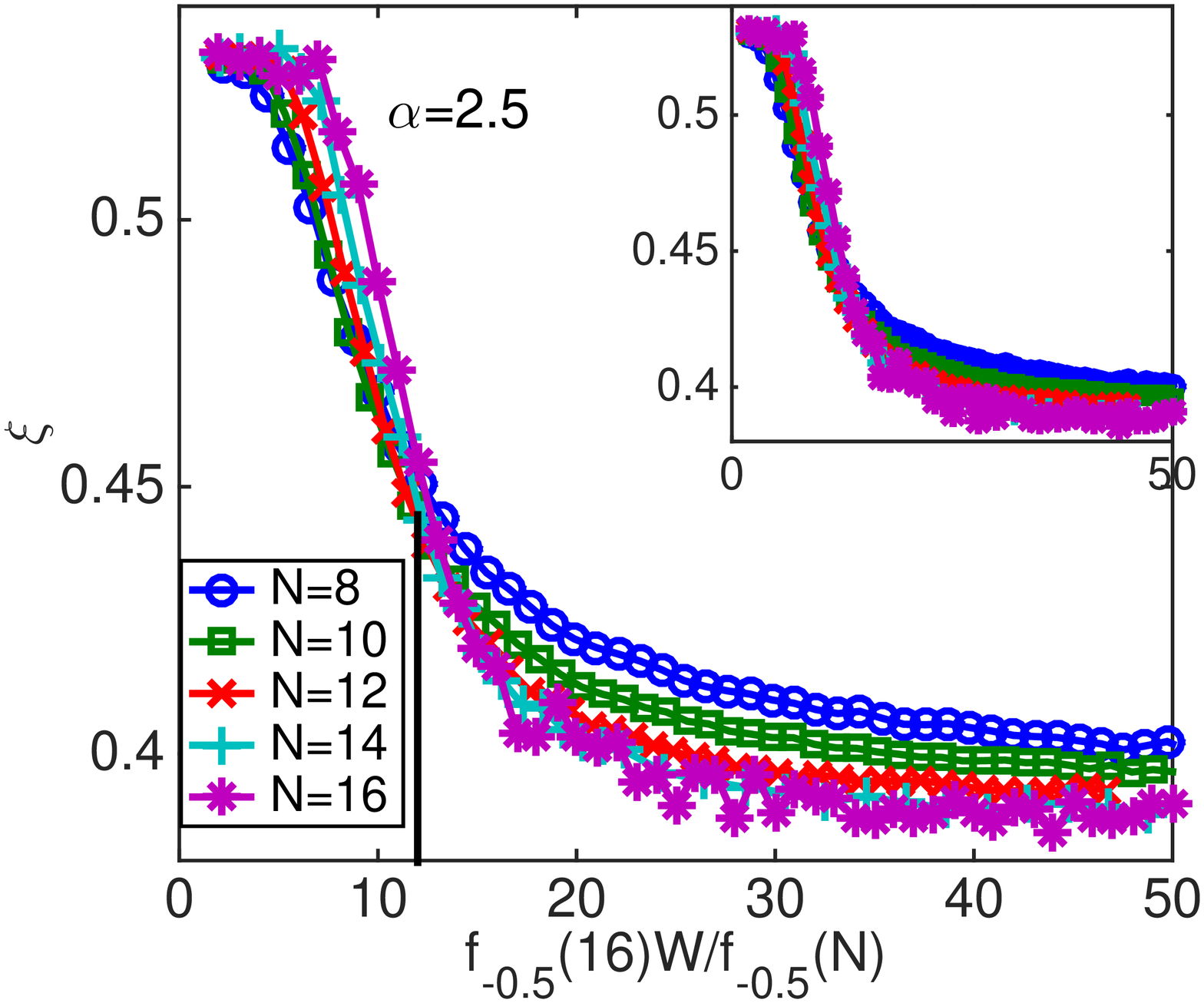}}\\
{\caption{\small  Rescaled  dependence of ergodicity (a) and level statistics (b) parameters on disordering $W$ compared to the original dependence (insets) for $R^{-2.5}$ spin-spin interactions..}
\label{fig:a2_5}}
\end{figure} 

\subsection{Results: Visual Inspection}
\label{sec:Results}

\begin{figure}[h!]
\centering
\subfloat[]{\includegraphics[scale=0.3]{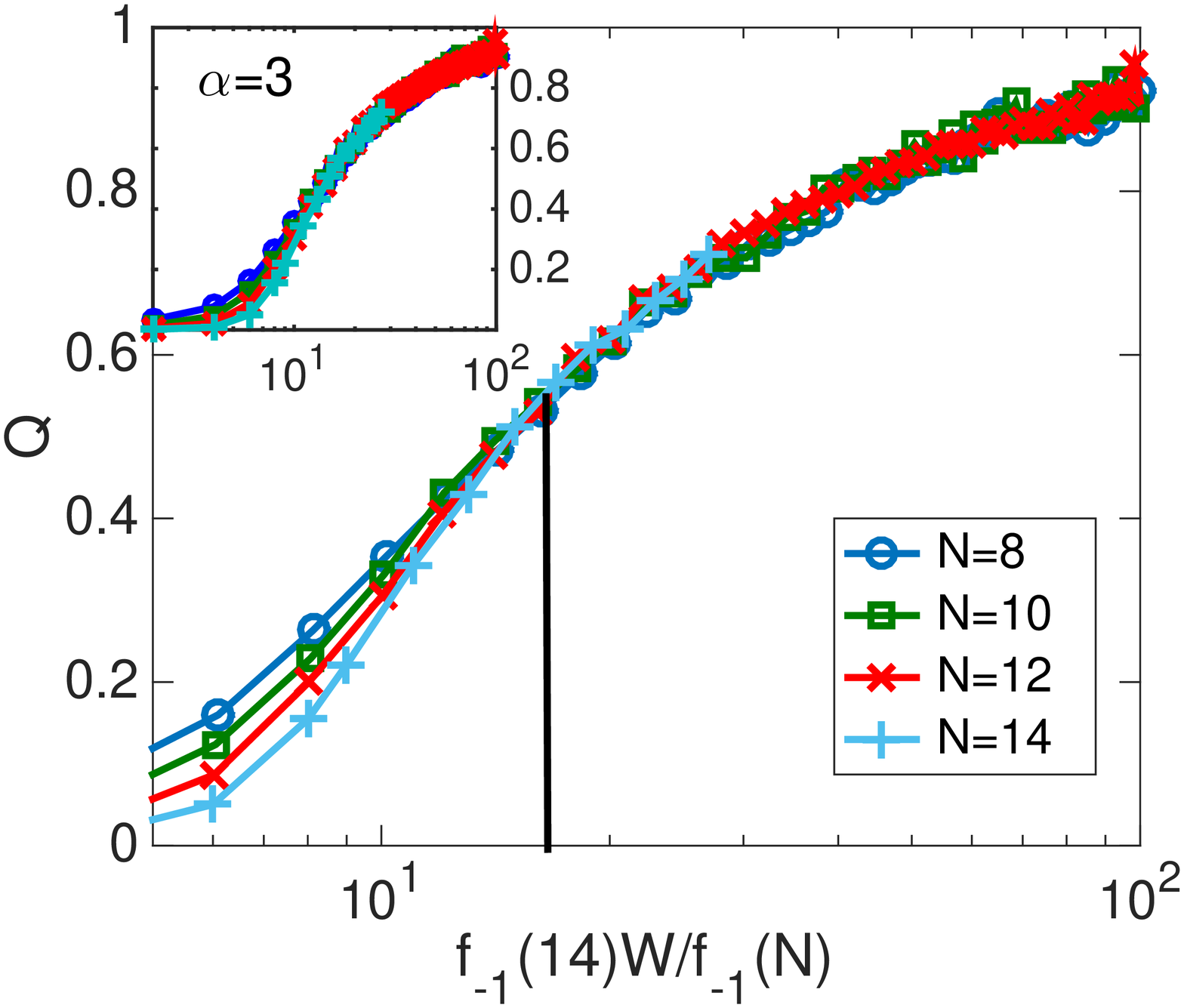}}\\
\subfloat[]{\includegraphics[scale=0.3]{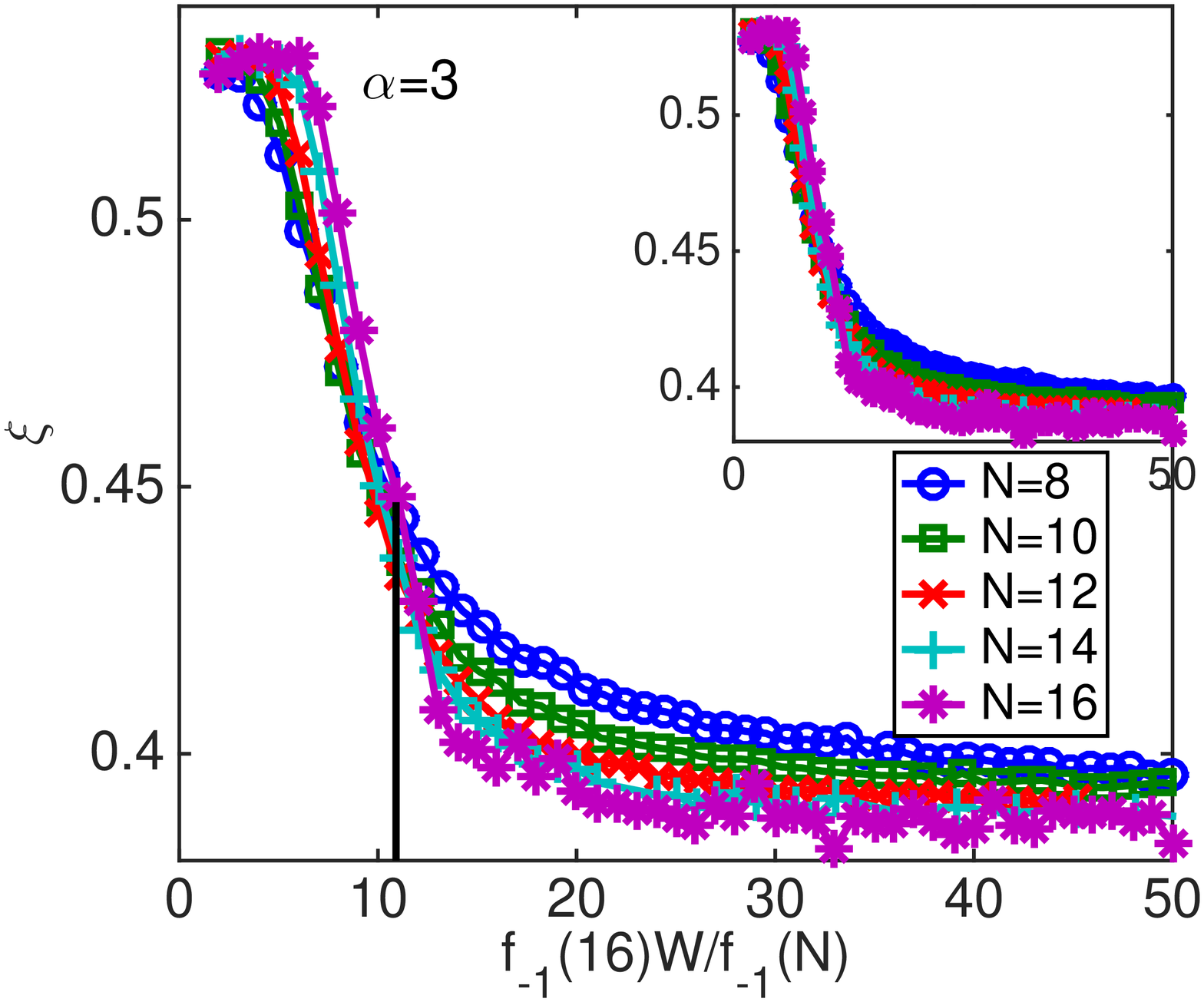}}\\
{\caption{\small  Rescaled  dependence of ergodicity (a) and level statistics (b) parameters on  disordering $W$ compared to the original dependence (insets) for $R^{-3}$ spin-spin interactions..}
\label{fig:a3}}
\end{figure}

In Figs. \ref{fig:a1}-\ref{fig:a3}, 
we show ergodicity (a) and level statistic (b) parameters versus disordering for different exponents $\alpha$. The main graphs show the dependence on the disordering rescaled using the functions $f_{a}(N)$ Eq. (\ref{eq:sc_func}) with the theoretically predicted parameters $a=2-\alpha$ (see Table \ref{tbl:scalinga}). The original dependencies are shown in the insets. 
All graphs show transitions between localization at strong disordering and delocalization at weak disordering. This is well confirmed by the level statistics behavior where the level statistics parameter approaches its Wigner Dyson limit ($\xi \approx 0.53$) at small disordering (see Ref. \cite{OganesyanParamStudy}).  The Poisson limit ($\xi \approx 0.38$) is also approached at large disordering and large system sizes $N>10$, while the deviations from the Poisson statistics at small sizes even in the classical spin limit $W\rightarrow \infty$ can be due to the finite size effects. 
However, size dependencies are very different at different exponents  showing almost no size dependence for $\alpha=3$ (Fig. \ref{fig:a3}) in agreement with Refs. \cite{PreprintML2,Pino} and the prominent shift of the transition towards large disordering in the case of a most slowly decaying interaction $\alpha=1$ (Fig. \ref{fig:a1}) in a qualitative agreement with the theory (Eq. (\ref{eq:MainScaling})) and the earlier numerical studies.\cite{PreprintML2}

The visual inspection of all graphs shows that the disordering rescaling leads to an approximate intersection of all rescaled curves at the same point. This observation is consistent with the expectations of the analytical theory and the intersection points can serve as estimates for the localization transition. Original graphs show the noticeable shift of transitions towards larger disordering with increasing the size at $\alpha \leq 1.75$, almost no displacement at $\alpha \geq 2.5$ and less conclusive behavior at intermediate exponents $\alpha =2, 2.25$. The results for $\alpha=1.5, 1.75$ can be more conclusive here than in Ref. \cite{PreprintML2} because we restrict the analysis to only nearly zero energy states that reduces the finite size effect in our consideration. These visual observations agree with the analytical theory predictions summarized in Table \ref{tbl:scalinga}. 

The transitions are getting narrower with increasing number of spins. This observation for the level statistics is consistent with the theoretical expectations of the instantaneous 
switch between  Wigner Dyson statistics and Poisson statistics at the localization transition point.\cite{ShklShapiro} Nearly discontinuous transition can also be expected for the ergodicity parameter in the case $\alpha<2d$ because according to the previous studies\cite{ab88Rv} (see also Sec. \ref{sec:Analyt}) the transition is caused by the small fraction of resonant pairs  at the transition point and the relative weight of this fraction $n_{p}(R_{*}) \sim N_{*}^{-\frac{\alpha-d}{d}}$ goes to zero with increasing the number of spins. Consequently one can neglect their effect on the spin-spin correlation function until the close vicinity of the transition is reached where resonant pairs begin ``talking to each other''. 

Using rescaled graphs one can also estimate the size dependent behavior of the transition points using the average position of the intersection of graphs for ergodicity parameters and level statistics. The estimates for the transition points are given in Table \ref{tbl:scaling}. Second and third lines there represent functional dependencies and asymptotic behaviors of critical disordering $W_{c}$ at large $N$. The estimate $W_{c} \approx 12.5U_{0}$ for the transition point in the  case $\alpha=3$ is consistent with the previous work \cite{PreprintML2,Pino}.  

\subsection{Quantitative Analysis}

The results of visual inspection cannot be taken as a strong evidence for the agreement of analytical theory and finite size scaling. To perform an independent estimate of the rescaling of localization transition with the system size we introduce the following procedure. 
Assume that the critical disordering increases with the system size as $W_{c}(N)$, while the width of the transition changes as $\Delta(N) \ll W(N)$  which agrees qualitatively with theoretical expectations and numerical results as discussed in Sec. \ref{sec:Results}. 

\begin{figure}[h!]
\centering
\includegraphics[scale=0.3]{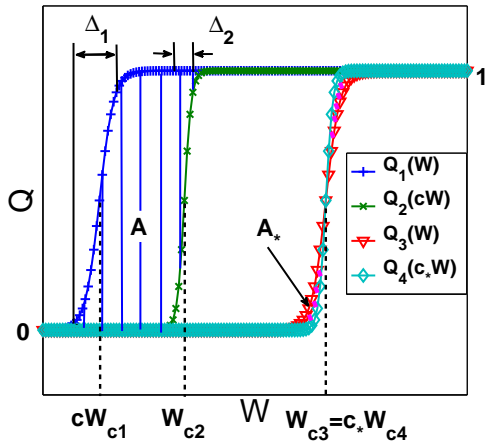}
{\caption{\small Schematic illustration of the optimum data rescaling occurring at the ratio of critical disorderings.}
\label{fig:ScalingIll}}
\end{figure}

Then for two data sets, say $Q_{1}(W, N_{1})$ and $Q_{2}(W, N_{2})$ ($N_{1}<N_{2}$), one can estimate the ratio of critical disorderings $c_{*}=W_{c}(N_{2})/W_{c}(N_{1})$ minimizing the squared deviation between the rescaled functions $Q_{1}(cW)$ and  $Q_{2}(W)$ defined as $\int_{0}^{\infty}dW(Q_{1}(cW)-Q_{2}(W))^2$  with respect to the parameter $c$. Practically these integrals are estimated as discrete sums $\sum_{i}(Q_{1}(cW_{i})-Q_{2}(W_{i}))^2$ using the discrete set $W_{i}=2, 4, ... 50$.   
As illustrated in Fig. \ref{fig:ScalingIll} the minimum should take place at $c\approx c_{*}$ because in that case the integral is determined  by the small  area $A_{*}$ comparable to the transition width $\Delta \ll W_{c}$ (functions $Q_{3}$ and $Q_{4}$) while for $c \neq c_{*}$ it is determined by the larger parameter of $W_{c}$ (functions $Q_{1}$ and $Q_{2}$). We restricted the consideration to the ergodicity parameter because the Poisson statistics limit is not realized properly for the small number of spins $N<12$ (see Figs. \ref{fig:a1}-\ref{fig:a3}.b) which overcomplicates data matching for level statistics. . 


\begin{figure}[h!]
\centering
\includegraphics[scale=0.3]{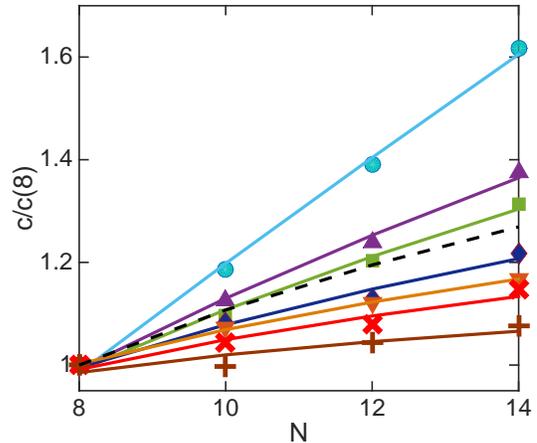}
{\caption{\small 
Rescaling constants vs. system size   for different interaction exponents $\alpha$ ($\circ$ for $\alpha=1$, $\bigtriangleup$  for $\alpha=1.5$, $\square$ for $\alpha=1.75$, $\diamond$ for $\alpha=2$, $\triangledown$ for $\alpha=2.25$, $\times$ for $\alpha=2.5$, and $+$ for $\alpha=3$. The lines show data fits by functions $f_{a}(N)$ Eq. (\ref{eq:sc_func}) optimized with respect to scaling exponents $a$ (see Table \ref{tbl:scaling}). Dashed black line shows the logarithmic dependence as a crossover between unlimited increase and saturation in the infinite size limit}
\label{fig:Scaling}}
\end{figure}

Using the Monte Carlo procedure developed earlier for fitting of multiple data sets \cite{abEcho} we determine the rescaling parameters $c_{N}$ for  $N=8, 10, 12$ matching of $Q_{N}(c(N)W)$ and $Q_{14}(W)$   as shown in Fig. \ref{fig:Scaling}. The obtained dependencies are compared to the logarithmic dependence (thick dotted line) serving as the crossover between the unlimited growing of $W_{c}(N)$ at small $\alpha$ and its saturation at $N\rightarrow \infty$ at large $\alpha$. The crossover line clearly separates the expected saturating $(\alpha>2)$  and unlimited growing $(\alpha < 2)$ regimes for $W_{c}(N)$ in agreement with the analytical expectations Eq. (\ref{eq:MainScaling}). 

We determine the optimum scaling exponent $a$ for the given interaction $R^{-\alpha}$ fitting the size dependence of the rescaling factors $c$ in Fig. \ref{fig:Scaling} with the function $f_{a}(N)$ Eq. (\ref{eq:sc_func}). The exponents $a$ determined for each $\alpha$ are summarized in Table \ref{tbl:scaling}. According to these estimates the critical disordering increases to infinity with increasing the system size for small interaction exponents $\alpha<2$ and saturates in the opposite case $\alpha>2$ in a full accord with the analytical theory. The results for the threshold case $\alpha=2$ are inconclusive. The obtained exponents are approximately consistent with the analytical theory predictions (see Eq. (\ref{eq:MainScaling}), fifth row of Table \ref{tbl:scaling}) at least for $\alpha<2$.  This justifies the use of the analytical theory to estimate the transition point $W_{c}$ in Sec. \ref{sec:Results}. 


\begin{table*}
 \caption{Scaling parameters for critical disordering determined using numerical and analytical methods for the long-range $R^{-\alpha}$ interactions.} 
 \label{tbl:scaling}
\centering
\scalebox{0.9}{
\begin{tabular}{|l|c|c|c|c|c|c|c|}
  \hline
  $\alpha$ & $1$ & $1.5$ &   $1.75$  & $2$  & $2.25$  & $2.75$   & $3$ \\
  \hline
     $W_{c}$ (estimate) & $3.67f_{1}(N)$ & $4.7f_{1/2}(N)$  & $6.06f_{1/4}(N)$  & $6.36f_{0}(N)$  & $6.2f_{-1/4}(N)$  & $7.2f_{-1/2}(N)$   & $7.6f_{-1}(N)$ \\ 
   \hline 
    $W_{c}$ (asymptotic) & $2.75N$ & $3.3N^{1/2}$  & $10.22N^{1/4}$  & $3.71\ln(N)$  & $28.3 - 29.2/N^{1/4}$  & $18.8 - 20/N^{1/2}$   & $12.5 - 30/N$ \\ 
  \hline
  $a$ ($Q$, $f_{a}(N)$) & $0.84$ & $0.4$   & $0.28$  & $-0.01$  & $-0.16$  & $-0.24$   & $-0.53$  \\
  \hline
$a$ (theory, $2d-\alpha$) & $1$ & $0.5$  & $0.25$  & $0$  & $-0.25$  & $-0.5$   & $-1$  \\ 
  \hline
  \end{tabular}
}
\end{table*}

\section{Conclusion}


In this paper we considered a many body localization problem in the system of interacting spins coupled by the long-range flip-flop and Ising interactions $R^{-\alpha}$. The absence of localization in the infinite size limit for the small exponents $\alpha<2d$ have been predicted. A critical disordering size dependence $W_{c} \propto N^{\frac{2d-\alpha}{d}}$ have been predicted and verified numerically (Table \ref{tbl:scalinga}) in a one dimensional system of interacting spins. 

Both visual inspection and quantitative analysis of localization transition size dependence are approximately consistent with the predictions of analytical theory for the threshold interaction exponent $\alpha_{c}=2d$ and the  critical disordering scaling. The transitions are getting narrower with increasing the system size and should probably approach the step function behavior in the infinite size limit in accord with qualitative expectations and numerical results. 

The predicted size dependence of the critical disordering can be tested in cold atomic systems varying number of atoms, interacting between them and disordering.\cite{PreprintML1,PreprintML2} Theory can also be applied to quantum two level systems in amorphous solids.\cite{AHVP} It was earlier suggested\cite{ab88Rv,abEcho} that the anomalously fast two level system relaxation observed there at very low temperature $T<30$mK can be due to the many-body interaction of two level systems. Indeed one can consider thermal two level systems with energies less or equal temperature as spins in our model. Then the energy disordering of such spins is given by the thermal energy $W \approx k_{B}T$, density of spins is determined as $n=P_{0}k_{B}T$ (here $P_{0} \approx 10^{43}$J$^{-1}$m$^{-3}$ is the typical two level system density of states\cite{abEcho}) and the spin-spin interaction at the average distance can be expressed as $\tilde{U} \approx k_{B}T P_{0}U_{0} \sim 10^{-3}k_{B}T$ where $U_{0}$ is a  $1/R^{3}$ interaction constant. Using these parameters one can estimate the critical system size needed for delocalization in a three dimensional system as $R_{*} \approx (P_{0}k_{B}TP_{0}U_{0})^{-\frac{1}{3}}$ (see Table \ref{tbl:scalinga}). At the temperature $T=20$mK this size can be estimated as $R_{*} \approx 7 \mu$m.   According to the present work  the interaction stimulated relaxation should disappear at smaller system sizes. Therefore it can be interesting to investigate two level system relaxation at temperatures $10-20$mK in ultrathin amorphous films with the thickness of few hundreds of $\AA$ngstr\"oms similar to those used in Josephson junction qubits\cite{Martinis} where no anomalous relaxation should be seen.




\section{Acknowledgement} 

This study has been stimulated by the recent work \cite{PreprintML2} investigating a many-body localization  with the long-range interaction. The author acknowledges Louisiana EPSCORE LA Sigma and LINK Programs for the support and Markus Muller and Kevin Osborn for useful discussions and suggestions.

\end{document}